\documentclass[epj]{svjour}

\usepackage{amsmath} 
\usepackage{amssymb} 
\usepackage[compatibility=false]{caption}
\usepackage{subcaption}
\usepackage{tabularx} 
\usepackage{pbox} 
\usepackage{float} 
\usepackage{wrapfig} 
\usepackage{graphicx}
\usepackage{dcolumn}
\usepackage{bm}
\usepackage{multirow} 
\usepackage{makecell} 
\usepackage[english]{babel}
\usepackage[utf8]{inputenc}  
\usepackage[T1]{fontenc}
\usepackage{url}
\usepackage{lineno}

\usepackage{siunitx}
\usepackage{upgreek}
\usepackage{cite}
\usepackage{microtype}

\usepackage{color}
\usepackage{xcolor}
\definecolor{bred}{rgb}{0.8, 0.0, 0.0}
\definecolor{pblue}{rgb}{0.2, 0.2, 0.6}

\usepackage[normalem]{ ulem }

\newcommand{\noyau}[2] {$^{\text{#1}}$#2}

\newcommand{\Stereo}{\textsc{Stereo}}

\begin{document}

\title{Improved STEREO simulation with a new gamma ray spectrum of excited gadolinium isotopes using FIFRELIN}


\author{H. Almaz\'an\inst{1}, L. Bernard\inst{2}, A. Blanchet\inst{3}, A. Bonhomme\inst{1,3}\thanks{Corresponding author: aurelie.bonhomme@mpi-hd.mpg.de}, C. Buck\inst{1}, A. Chebboubi\inst{6}, P. del Amo Sanchez\inst{4}, I.~El~Atmani\inst{3}\thanks{Present address: Hassan II University, Faculty of Sciences, A\"in Chock, BP 5366 Maarif, Casablanca 20100, Morocco}, J. Haser\inst{1}, F. Kandzia\inst{5}, S. Kox\inst{2}, L. Labit\inst{4}, J. Lamblin\inst{2}, A. Letourneau\inst{3}, D. Lhuillier\inst{3}, M. Lindner\inst{1}, O. Litaize\inst{6}, T. Materna\inst{3}, A. Minotti\inst{3}\thanks{Present address: Univ.~Grenoble Alpes, Universit\'e Savoie Mont Blanc, CNRS/IN2P3, LAPP, 74000 Annecy, France}, H. Pessard\inst{4}, J.-S. R\'eal\inst{2}, C. Roca\inst{1}, T. Salagnac\inst{2}\thanks{Present address: IPNL, CNRS/IN2P3, Univ.~Lyon, Université Lyon 1, 69622 Villeurbanne, France}, V. Savu\inst{3}, S. Schoppmann\inst{1}, V. Sergeyeva\inst{4}\thanks{Present address: IPN Orsay, CNRS/IN2P3, 15 rue Georges Clemenceau, 91406 Orsay, France}, T. Soldner\inst{5}, A. Stutz\inst{2}, L. Thulliez\inst{3} \and M. Vialat\inst{2}}

%
%
\institute{Max-Planck-Institut f\"ur Kernphysik, Saupfercheckweg 1, 69117 Heidelberg, Germany \and Univ.~Grenoble Alpes, CNRS, Grenoble INP, LPSC-IN2P3, 38000 Grenoble, France \and IRFU, CEA, Universit\'e Paris-Saclay, 91191 Gif-sur-Yvette, France \and Univ.~Grenoble Alpes, Universit\'e Savoie Mont Blanc, CNRS/IN2P3, LAPP, 74000 Annecy, France \and Institut Laue-Langevin, CS 20156, 38042 Grenoble Cedex 9, France \and CEA, DEN, DER, SPRC, F-13108 Saint Paul Lez Durance, France}
\date{}
%

\abstract{
The {\textsc{Stereo}} experiment measures the electron antineutrino spectrum emitted in a research reactor using the inverse beta decay reaction on H nuclei in a gadolinium loaded liquid scintillator. The detection is based on a signal coincidence of a prompt positron and a delayed neutron capture event.
The simulated response of the neutron capture on gadolinium is crucial for the comparison with data, in particular in the case of the detection efficiency.
Among all stable isotopes, $^{155}$Gd and $^{157}$Gd have the highest cross sections for thermal neutron capture. The excited nuclei after the neutron capture emit gamma rays with a total energy of about 8\,MeV. The complex level schemes of $^{156}$Gd and $^{158}$Gd are a challenge for the modeling and prediction of the deexcitation spectrum, especially for compact detectors where gamma rays can escape the active volume. With a new description of the Gd(n,~$\gamma$) cascades obtained using the \textsc{Fifrelin} code, the agreement between simulation and measurements with a neutron calibration source was significantly improved in the \Stereo~experiment. A database of ten millions of deexcitation cascades for each isotope has been generated and is now available for the user.
} 
\authorrunning{H.~Almaz\'an et al.}
\titlerunning{Improved STEREO simulation with a $\gamma$-ray spectrum of excited Gd isotopes using FIFRELIN}
\maketitle

The {\textsc{Stereo}} experiment~\cite{Allemandou:2018vwb} detects electron antineutrinos produced in a compact research reactor at the Institut Laue-Langevin (ILL) in Grenoble, France. With the \Stereo~data, it is possible to test the hypothetical existence of a sterile neutrino in the eV mass range~\cite{Almazan:2018wln}. Antineutrinos are detected via the inverse beta decay reaction (IBD) $\bar{\nu_e}+p \rightarrow e^+ +n$ in a gadolinium (Gd) loaded organic liquid scintillator (LS)~\cite{Buck:2018cac}. The positron ionization gives rise to a prompt signal --~related to the antineutrino energy~--, while the neutron thermalizes and diffuses in the liquid. The gamma emission following the radiative capture of the neutron creates a delayed signal after typical coincidence time of a few tens of $\upmu$s.

The capture time can be strongly reduced by the addition of Gd with its very high cross section for thermal neutron capture ($\sim 10^5$\,barn for some of the isotopes) to the LS. At a Gd-concentration of 0.2\,wt.\% in {\textsc{Stereo}}, the average capture time is $\SI{18}{\micro\second}$, one order of magnitude lower than for an unloaded scintillator where capture is mainly by hydrogen (H). More than 80\% of the neutrons are captured on Gd. The resulting excited Gd-nuclei decay to the ground state by emitting on average four gammas with a total energy of about 8\,MeV. This is well above the typical energies of the natural radioactivity backgrounds and the 2.2\,MeV line of H(n,~$\gamma$), providing a clean detection channel. Therefore, several past, present, and upcoming neutrino detectors take advantage of the Gd-loaded LS technology. All of these experiments rely on a precise knowledge of the Gd-deexcitation process, where the gamma multiplicity and single energies are especially relevant for segmented and/or compact detectors such as {\textsc{Stereo}}. These quantities play a crucial role in the reconstructed spectrum of the experiments, since high energy gammas are more likely to escape the detector or leave only a part of their energy via Compton scattering, and populate thereby the tail of the reconstructed peak. 
The DANSS experiment, for example, reported some tension between calibration and simulation data in the low energy part of the Gd spectrum~\cite{Alekseev:2018efk}. To determine the neutron detection efficiency in the Daya Bay experiment, four different models are used to estimate the gamma energy and multiplicity distributions~\cite{Adey:2018qct}. An accurate modeling of these distributions is also of major relevance in water Cherenkov detectors with Gd-loading~\cite{jparc}, since the light production is very sensitive to the energies of the single gammas due to the Cherenkov threshold. Therefore, a good understanding of the cascade for the most relevant isotopes \noyau{156}{Gd} and \noyau{158}{Gd} is of primary importance to reduce the systematic uncertainties. However, none of these isotopes have complete experimentally known nuclear level schemes and branching ratios up to 8\,MeV.
The cause of small discrepancies between the {\textsc{Stereo}} data and simulation was identified to be largely due to an inaccurate description of the gamma deexcitation of the Gd nuclei. By using \noyau{156}{Gd}$^*$ and \noyau{158}{Gd}$^*$ gamma cascades from the nucleus deexcitation code \textsc{Fifrelin} developed at CEA-Cadarache (France) these discrepancies are reduced.


The \textsc{Fifrelin} code is developed for the evaluation of fission data and has already proven its ability to make accurate predictions regarding neutron and gamma properties~\cite{Litaize2013,Litaize2015,Litaize2018}. It provides important information on crucial parameters like the gamma multiplicity and the particle energies. The code is made of two parts: one is the assignment of the initial state of the fission fragment and the other is the deexcitation process. For the \Stereo~experiment, only the deexcitation part is used. Once the initial states are given, a set of nucleus level schemes is sampled allowing to take into account nuclear structure uncertainties. The deexcitation processes are then performed within a Monte-Carlo Hauser-Feschbach framework, based on Be\v{c}v\'{a}\v{r}'s algorithm \cite{Becvar1998}, and extended to the n/$\gamma$ emission by R\'{e}gnier \cite{Fifrelin2016}. 

\begin{figure}[H]  
    \centering 
    \includegraphics[scale=0.3]{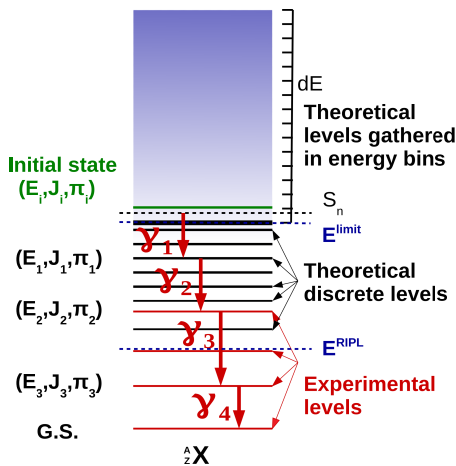} 
    \caption{\textit{Sketch of a $\gamma$-cascade simulated in \textsc{Fifrelin}.}} 
    \label{fig:Fifrelin_deexcitationScheme} 
\end{figure} 

After a thermal neutron capture (E$_n$=25\,meV), the \noyau{156}{Gd*} and \noyau{158}{Gd*} nuclei have an excitation energy approximately equal to the compound nucleus neutron separation energy S$_n$ (E$_n\,\ll\,$S$_n$), equal to 8.536\,MeV and 7.937\,MeV, respectively. Knowing the \noyau{155}{Gd}, \noyau{157}{Gd} ground state spin-parity J$^{\pi}$=3/2$^{-}$, the selection rules give to the excited nuclei a parity -1 and two allowed spins of 1$\hbar$ and 2$\hbar$. Following the latest nuclear data evaluations~\cite{Brown20181short, jendl, Mughabghab} a 2$\hbar$ spin is assigned to both nuclei, corresponding to their first resonance spin.
In \textsc{Fifrelin}, a realization of the nuclear level scheme (fig.~\ref{fig:Fifrelin_deexcitationScheme}) uses all the experimental knowledge and the missing information comes from nuclear models:

\begin{itemize}\label{enum:levelSchemeBuilding}
	\item for E~$\leq\text{E}^{\text{RIPL}}$, all the energy levels are known and are retrieved from the Reference Input Parameter Library (RIPL-3) database \cite{RIPL3}. If a spin and/or a parity are missing, the code samples them from  the momentum and parity theoretical distributions detailed below. 
	\item for $\text{E}^{\text{RIPL}}<$~E~$\leq\text{E}^{\text{limit}}$, only a few levels are experimentally determined. Additional discrete levels are then sampled until the level number matches the theoretical level density. This is done until $\text{E}^{\text{limit}}$, defined by a nuclear level density set to 5$\cdot$10$^{4}$ MeV$^{-1}$ (default value).
	\item for E~>~$\text{E}^{\text{limit}}$, the number of levels is innumerable and corresponds to the continuum. Therefore, levels are gathered in energy bins (dE=10 keV by default) having a specific $J^{\pi}$ given by a model.
\end{itemize}
\noindent
The values of E$^{\text{RIPL}}$, E$^{\text{limit}}$ and E$^{\text{M}}$ for the relevant Gd isotopes are given in Table \ref{tab:Gd_Fifrelin_Energies}.
The theoretical nuclear level density $\rho (\text{E,J,}\pi)$ used to complete the level scheme is

\begin{equation}\label{eq:NLD}
\rho (\text{E,J,}\pi)=\rho_{\text{tot}}(\text{E})~\text{P(J|E)~P(}\pi)
\end{equation}
\noindent
with $\rho_{\text{tot}}(\text{E})$ the total nuclear level density, P(J|E) the energy dependent angular momentum distribution and P($\pi$) the parity distribution. The positive and negative parities are assumed to be equiprobable, P(\,$\pi$\,=\,$\pm$\,1\,)\,=\,0.5.
The angular momentum distribution is
\begin{equation}\label{eq:angularMomentumDistribution}
\text{P(J|E)}=\frac{\text{2J+1}}{2\sigma^{2}(\text{E})}\text{exp}\left(-\frac{(\text{J+1/2})^{2}}{2\sigma^{2}(\text{E})} \right)
\end{equation}
\noindent 
with $\sigma^{2}$(E) the spin cut-off parameter defining the dispersion of the nucleus angular momentum. More information on its origin and its parametrization can be found in~\cite{Bethe1936, RIPL3}. The total nuclear level density follows the Composite Gilbert Cameron Model (CGCM)~\cite{GilbertCameron1965}, using the Constant Temperature Model (CTM) at low energy and the Fermi Gas Model (FGM) at high energy

\begin{equation} \label{eq:levelDensity_CGCM}
\rho_{\text{tot}}^{\text{CGCM}}(\text{E}) = \left\{
\begin{array}{ll}
  \rho_{\text{tot}}^{\text{CTM}}(\text{E})	 &     \text{, for } \text{E} \leq \text{E}_{M}\\
  \rho_{\text{tot}}^{\text{FGM}}(\text{E})	 &  \text{, for } \text{E} > \text{E}_{M}\\
\end{array}
\right.
\end{equation}
\noindent
with $\text{E}_M$ the energy where the CTM and FGM nuclear level densities match along with their derivatives. The CGCM parametrizations used here can be found in \cite{RIPL3}.

\begin{table}[H]
	\centering
	\caption{\textit{Threshold energies for \noyau{156}{Gd} and \noyau{158}{Gd} used in \textsc{FIFRELIN}.} \label{tab:Gd_Fifrelin_Energies}}
	\begin{tabular}{c | c c c }
		& $\text{E}^{\text{RIPL}}$ (MeV) & $\text{E}^{\text{limit}}$ (MeV) & $\text{E}^{\text{M}}$ (MeV) \\
		\hline
		\noyau{156}{Gd} & 1.366 & 5.306 & 6.605 \\
		\noyau{158}{Gd} & 1.499 & 5.402 & 6.376 \\
	\end{tabular}
\end{table}

During a deexcitation step, all the transition probabilities $\Gamma_{p}(i\rightarrow f, \alpha)$ to go from a given initial state ($i$) to a final state ($f$) in emitting a particle $p$ with given properties ($\alpha$) are computed. Then, one transition is sampled among all of them.
Generally, models only give access to the average partial width $\overline{\Gamma}_{p}(I\rightarrow F, \alpha)$ associated to a transition from an initial set of levels ($I\equiv$[E, J$^{\pi}$]$_i$) to a final set ($F\equiv$[E, J$^{\pi}$]$_f$), having both a given $J^{\pi}$:

\begin{equation}\label{eq:averagePartialWidth}
\overline{\Gamma}_{p}(I\rightarrow F, \alpha) = \Bigg \langle\sum_{f\in F} \Gamma_{p}(i\rightarrow f, \alpha)\Bigg \rangle _{i\in I}.
\end{equation}
\noindent
Finally, the partial width to go from $I$ to $F$ is

\begin{equation}
\Gamma_{p}(I\rightarrow F, \alpha) = \overline{\Gamma}_{p}(\epsilon_p, \alpha)\delta (\alpha,\text{J}^{\pi}_i,\text{J}^{\pi}_f) y(I\rightarrow F)\rho (\text{E}_f,\text{J}^{\pi}_f)\text{dE}
\end{equation}
\noindent
where $\epsilon_p$ is the transition energy, and with $\delta (\alpha,\text{J}^{\pi}_i,\text{J}^{\pi}_f)$ accounting for spin and parity selection rules, $y(I\rightarrow F)$ the Porter-Thomas factor simulating the transition probability fluctuation described by a $\chi^{2}$ distribution with [$\rho(\text{E}_i,\text{J}^{\pi}_i)$dE$\times \rho(\text{E}_f,\text{J}^{\pi}_f)$dE] degrees of freedom \cite{Porter-Thomas1956} and\\$\rho (\text{E}_f,\text{J}^{\pi}_f)\text{dE}$ the number of levels in $F$.  
\newline
\indent
At $S_n$, neutron emission is unlikely. Conversion electrons are taken into account using BrIcc code~\cite{Kibedi2008}. 
For gamma emission, the average partial width ($\overline{\Gamma}_{\gamma}(\epsilon_{\gamma},\text{XL})$) depends on the emitted gamma energy ($\epsilon_{\gamma}$), its type X (electric or magnetic), its multipolarity L and the radiative strength function model ($f_{\text{XL}}$):

\begin{equation}
\overline{\Gamma}_{\gamma}(\epsilon_{\gamma},\text{XL})=\epsilon_{\gamma}^{\text{2L+1}}f_{\text{XL}}(\epsilon_{\gamma})/\rho(\text{E}_i,\text{J}^{\pi}_i)
\end{equation}
\noindent 
The E1 transition is described by the Enhanced Generalized Lorentzian Model (EGLO) which is a Lorentzian function where an asymptotic term is added to better reproduce low energy experimental data \cite{KopeckyUhl,RIPL3}. The other XL transitions are best described by a Standard Lorentzian Model (SLO)~\cite{Brink1957} defined by:

\begin{equation}
f_{\text{XL}}(\epsilon_{\gamma}) = \frac{f_{\text{E1}}^{\text{EGLO}}(B_n)}{R}\frac{f_{\text{XL}}^{\text{SLO}}(\epsilon_{\gamma})}{f_{\text{XL}}^{\text{SLO}}(B_n)}
\end{equation} 
\noindent
where $B_n$ is the neutron binding energy and $R$ a nucleus mass dependent ratio. More details can be found in~\cite{RIPL3}.

\begin{figure}[H]  
	\centering 
	\includegraphics[width=0.5\textwidth]{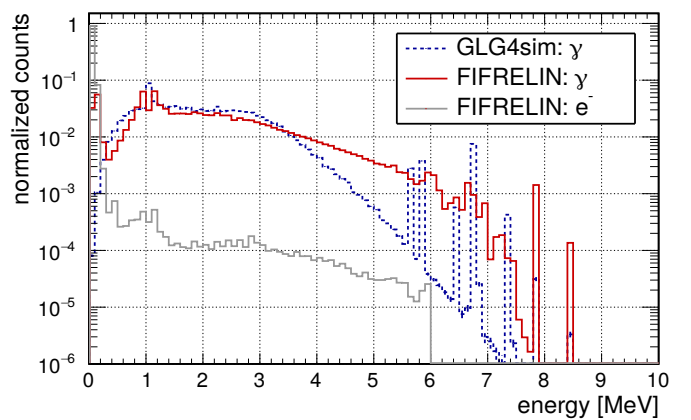} 
	\caption{\textit{Energy distribution of the cascade deexcitation products: the GLG4sim simulation provides only gammas (dashed blue), whereas both gammas (red) and conversion electrons (gray) are generated in the \textsc{Fifrelin} simulation.}} 
	\label{fig:sum_singles} 
\end{figure}

The \Stereo~simulation is based on \textsc{Geant4} libraries \cite{GEANT4} and includes the detailed geometry of the detector, the description of its response with special emphasis on light emission and collection \cite{Allemandou:2018vwb}.
Neutron transportation is handled by the \texttt{NeutronHP} libraries, in which microscopic interaction cross-sections are from the ENDF/B-VII.1 evaluation.
Standard deexcitation processes in \textsc{Geant4} do not offer a satisfactory treatment on an event-by-event basis regarding energy conservation. As a consequence, when a neutron is captured on a Gd isotope, standard \texttt{NeutronHP} processes are bypassed and an user-defined process is used in the \Stereo~simulation. An empirical gamma-cascade treatment was initially performed using an additional support for the GLG4sim package \cite{GLG4sim_Gd}, developed specifically for neutrino detection in LS. This implementation gave satisfactory results in larger detectors for well contained energy depositions.
In the new \Stereo~simulation, the deexcitation cascades from \textsc{Fifrelin} are directly used. In both cases, the deexcitation products are generated isotropically.
For natural Gd, the gamma multiplicity per cascade is about 4, and differs only by a few percents between GLG4sim and \textsc{Fifrelin}. The major difference between the codes can be seen in the energy distribution of single deexcitation products, presented in fig.~\ref{fig:sum_singles}. About 15\,\% of the gammas generated in the \textsc{Fifrelin} simulation have an energy higher than 3.5\,MeV, while they only account for 7\,\% in the GLG4sim modeling.
Recently, independent measurements~\cite{jparc} have shown that these high-energy gammas are needed for an accurate description of the cascade.
This is of primary importance for the \Stereo~detector since at such energies around 5\,MeV, the mean free path of a gamma in the LS is 40\,cm, comparable to the characteristic size of a \Stereo~cell. Conversion electrons are present in about 70\,\% of the cascades with a most probable energy of 70\,keV. Due to very low emission probability, electrons of more than 200\,keV represent less than 1\,\% of the sample.

\begin{figure}
	\centering
	\includegraphics[width=0.5\textwidth]{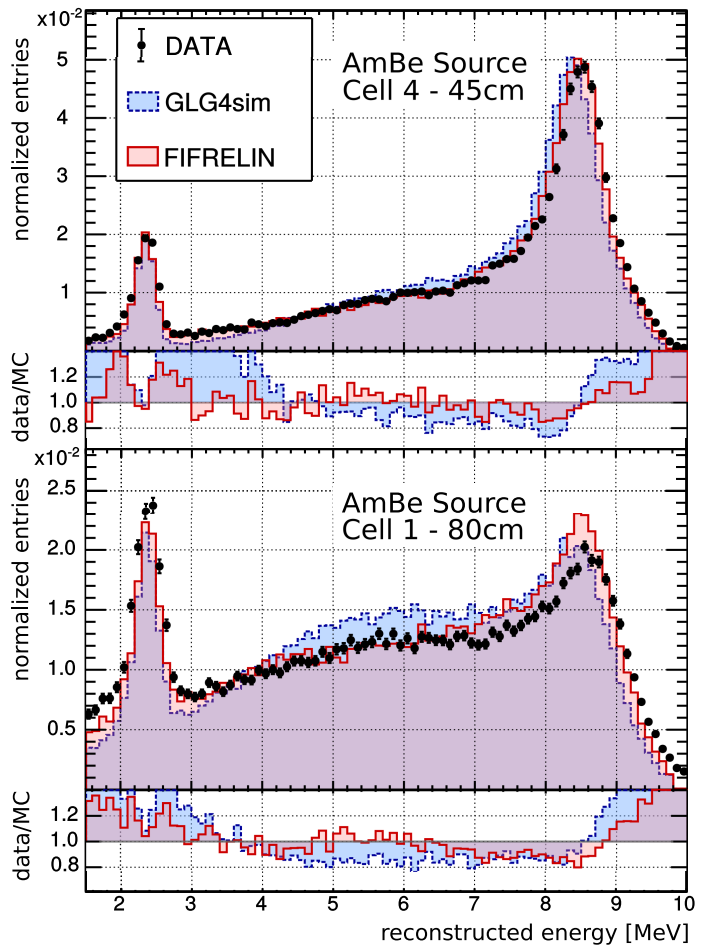}
	\caption{\textit{Reconstructed energy spectra from neutron captures from an AmBe source in a central (upper plot) and a top position (lower plot). Data points are in black, and GLG4sim (\textsc{Fifrelin}) simulation is in blue (red).}\label{fig:n_spectrum}} 
\end{figure}

In the \Stereo~experiment, the neutron response -- characterizing the delayed signal -- is monitored using an americium-beryllium (AmBe) source deployed regularly in 5 of the 6 identical 91\,cm high target cells at 5 different vertical levels (10, 30, 45, 60 and 80\,cm from the bottom). Neutrons are produced at a $15\cdot 10^3\,\SI{}{\per\second}$
rate through the reaction: $\alpha + ^9\text{Be} \rightarrow ^{12}\text{C} + \text{n}$. In about 60\% of the cases \cite{SCHERZINGER201574}, the neutron emission is accompanied by a 4.4\,MeV gamma from carbon deexcitation. A coincidence selection is then applied to isolate the neutrons from these gammas and to get a clean and pure neutron capture sample without background: delayed signals are searched in a time window of $\SI{100}{\micro\second}$ after a prompt 4.4\,MeV gamma signal, and contributions from random coincidences (accidental background) are statistically subtracted. The resulting delayed energy spectrum is presented in fig.~\ref{fig:n_spectrum}, where both the 2.2\,MeV peak from H(n,~$\gamma$) and the $\sim$\,8\,MeV from Gd(n,~$\gamma$) are visible, along with the simulations. The shape of the tail over all the energy range is greatly improved with the \textsc{Fifrelin} description, for central positions, as well as for border positions, more sensitive to escaping gammas.
Performing a Kolmogorov-Smirnov test on the tail from 3 to 7 MeV, we find an agreement with a probability of 11\%, providing no indication for a systematic effect in the description at the central position, whereas the test showed clear incompatibility between data and the GLG4sim spectrum at more than 5 standard deviations.  As expected, the presence of higher energy gammas tends to correct the balance between low and high energy events in the tail.
The \Stereo~energy scale being anchored to the low energy gamma of \noyau{54}{Mn}~\cite{Allemandou:2018vwb}, the mean positions of the reconstructed peaks are artificially higher than literature values, due to quenching effects. These non-linearity effects are calibrated and taken into account in the \textsc{Stereo} simulation. In order to assess the improvement of this new simulation without considerations on any residual linear systematic effect on the energy scale, the reconstructed energy of both simulations is scaled such that the mean position of the H(n,~$\gamma$) peak from the simulation matches the data. In this way, the agreement for the Gd(n,~$\gamma$) peak is evaluated relatively to the H capture peak at 2.2\,MeV, and an agreement for the Gd mean peak position at the sub-percent level is achieved with the \textsc{Fifrelin} simulation.

The neutrino detection uncertainty in {\textsc{Stereo}} is dominated by the systematic uncertainty on the delayed neutron detection efficiency. Beyond selection cuts related to the event topology~\cite{Almazan:2018wln}, the delayed event of an IBD candidate is required in this article to be within a time window of (0.5--70)\,$\upmu$s after the prompt signal. The largest inefficiency is coming from the (4.5\,--\,10)\,MeV energy cut on the delayed event set in order to select only Gd events. The lower threshold is chosen to maximize the signal-to-background ratio and to minimize the systematic uncertainties, and it is clear from fig.~\ref{fig:n_spectrum} that a significant part of the Gd events -- with large energy leakage -- is not included. The correct description of the spectrum over the full energy range is therefore essential. 

To quantify the impact of the event selection cuts, the neutron capture spectra are divided in two parts: a H window (1.5\,--\,3)\,MeV and a Gd window (3\,--\,10)\,MeV.
The ratio of events in the Gd window (N$_{\text{Gd}}$) and the total sum (N$_{\text{Gd}}$+N$_{\text{H}}$) is defined as the Gd-fraction ($\varepsilon_{\text{Gd}}$):
\begin{equation}
	\varepsilon_{\text{Gd}} = \frac{\text{N}_{\text{Gd}}}{\text{N}_{\text{H}}+\text{N}_{\text{Gd}}}
\end{equation}
The impact on the delayed selection cuts (time and energy) are evaluated by defining the IBD efficiency $\varepsilon_{\text{IBD}}$, fraction of events in the (N$_{\text{Gd}}$) passing the tighter delayed selection used in the neutrino analysis:
\begin{equation}
	\varepsilon_{\text{IBD}} = \frac{\text{N}_{(4.5\,-\,10)\,\text{MeV}~\text{and}~(0.5\,\upmu\text{s} \,<\Delta\text{T}\,<\,70\,\upmu\text{s})}}{\text{N}_{\text{Gd}}}
\end{equation}
The total delayed detection efficiency $\varepsilon_{\text{tot}}$ is then the product of these two terms:
\begin{equation}
	\varepsilon_{\text{tot}}= \varepsilon_{\text{Gd}}\cdot\varepsilon_{\text{IBD}}\\
\end{equation} 

The numbers in Table~\ref{tab:data_mc_efficiency} illustrate that the {\textsc{Stereo}} data favor the Gd spectrum from the \textsc{Fifrelin} events as compared to the GLG4sim simulation. Using GLG4sim, the ratio R=$\varepsilon_{\text{tot}}^{\text{Data}}$/$\varepsilon_{\text{tot}}^{\text{MC}}$ quantifying the agreement between data and MC for the total efficiency was found to be 0.9537 at the most central calibration point as shown in Table~\ref{tab:data_mc_efficiency}. In the new simulation, $\varepsilon_{\text{tot}}^{\text{MC}}$ matches the data within 1\% (R=0.9953).
For $\varepsilon_{\text{Gd}}$, small discrepancies remain, mainly in the border positions. The data/MC ratio of $\varepsilon_{\text{Gd}}$ is very sensitive to the treatment of the neutron propagation. In particular the modeling of neutron scattering and thermal diffusion in the detector as well as the neutron detection cross-section could induce an additional mismatch. The border regions are more sensitive to such inaccuracies. Therefore, as an extreme case, the calibration data at the top of cell 1 was investigated, for which the source was located only 12\,cm (8\,cm) from the target wall (cell top).
Overall very good data/MC agreement is achieved for $\varepsilon_{\text{IBD}}$ due to the improvements related to the new simulation input of \textsc{Fifrelin} (see fig.~\ref{fig:n_spectrum}).


\begin{table}[H]
\caption{\textit{Ratios Data/MC of the partial ($\varepsilon_{\text{Gd}}$, $\varepsilon_{\text{IBD}}$) and total efficiencies for GLG4sim and \textsc{Fifrelin} simulations, in the case of the deployment of the AmBe cell at a central position (second column) and at a border position (third column). Only statistical uncertainties are quoted.} \label{tab:data_mc_efficiency}}
	\begin{tabular}{c c | c | c }
		& & Cell 4 (central) & Cell 1 (border) \\
		& & Central position & Top position\\
		\hline
		\multirow{ 2}{*}{$\varepsilon_{\text{Gd}}^{\text{Data}}$/$\varepsilon_{\text{Gd}}^{\text{MC}}$} & GLG4sim & $0.9744\pm0.0003$ & $0.9436\pm0.0013$ \\
		 & \textsc{Fifrelin} & $0.9918\pm0.0003$ & $0.9682\pm0.0013$\\
		 \hline
		 \multirow{ 2}{*}{$\varepsilon_{\text{IBD}}^{\text{Data}}$/$\varepsilon_{\text{IBD}}^{\text{MC}}$} & GLG4sim & $0.9814\pm0.0004$ & $0.9957\pm0.0018$ \\
		 & \textsc{Fifrelin} & $1.0035\pm0.0005$ & $1.0091\pm0.0019$\\
		 \hline
		 \hline
		 \multirow{ 2}{*}{$\varepsilon_{\text{tot}}^{\text{Data}}$/$\varepsilon_{\text{tot}}^{\text{MC}}$} & GLG4sim & $0.9562\pm0.0005$ & $0.9396\pm0.0025$ \\
		 & \textsc{Fifrelin} & $0.9953\pm0.0006$ & $0.9770\pm0.0022$\\
	\end{tabular}
\end{table}   

In summary, the correct description of the deexcitation process of the Gd nuclei after neutron capture is crucial for neutrino detection experiments using Gd in general. In particular, this is the case for small detectors sensitive to gamma escape such as \Stereo. Since nuclear level schemes are not completely experimentally known, nuclear models are needed. The \Stereo~description of the Gd cascade was greatly improved using the \textsc{Fifrelin} nuclear code, making benefit of the most updated nuclear databases and user feedback on nuclear evaluations. We make available ten millions of deexcitation cascades for each isotope~\cite{fifrelinDATA}, since other running and upcoming projects might profit from these data as well. 

\acknowledgement
This work is supported by the French National Research
Agency (ANR) within the Project No. ANR-13-BS05-0007
and the programs P2IO LabEx (ANR-10-LABX-0038) and
ENIGMASS LabEx (ANR-11-LABX-0012). We acknowledge the support of the CEA, CNRS/IN2P3, the ILL and the Max-Planck-Gesellschaft.

\bibliographystyle{ieeetr}
\bibliography{mybibliography}

\begin{thebibliography}{10}

\bibitem{Allemandou:2018vwb}
N.~Allemandou {\em et~al.}, ``{The \textsc{Stereo} Experiment},'' {\em JINST},
  vol.~13, no.~07, p.~P07009, 2018.

\bibitem{Almazan:2018wln}
H.~Almazán {\em et~al.}, ``{Sterile Neutrino Constraints from the
  \textsc{Stereo} Experiment with 66 Days of Reactor-On Data},'' {\em Phys.
  Rev. Lett.}, vol.~121, no.~16, p.~161801, 2018.

\bibitem{Buck:2018cac}
C.~Buck, B.~Gramlich, M.~Lindner, C.~Roca, and S.~Schoppmann, ``{Production and
  Properties of the Liquid Scintillators used in the \textsc{Stereo} Reactor
  Neutrino Experiment},'' {\em JINST}, vol.~14, no.~01, p.~P01027, 2019.

\bibitem{Alekseev:2018efk}
I.~Alekseev {\em et~al.}, ``{Search for sterile neutrinos at the DANSS
  experiment},'' {\em Phys. Lett. B7}, vol.~87, pp.~56--63, 2018.

\bibitem{Adey:2018qct}
D.~Adey {\em et~al.}, ``{Improved Measurement of the Reactor Antineutrino Flux
  at Daya Bay},'' 2018\\ arXiv:1808.10836.

\bibitem{jparc}
K.~Hagiwara {\em et~al.}, ``{Gamma Ray Spectrum from Thermal Neutron Capture on
  Gadolinium-157},'' {\em PTEP}, vol.~2019, no.~2, p.~023D01, 2019.

\bibitem{Litaize2013}
O.~Litaize~et al., ``{Investigation of n+$^{238}${U} Fission Observables},''
  {\em Nuclear Data Sheets}, vol.~118, pp.~216--219, 2014.

\bibitem{Litaize2015}
O.~Litaize~et al., ``{Fission modelling with FIFRELIN},'' {\em Eur. Phys. J.
  A}, vol.~51, no.~177, pp.~1--14, 2015.

\bibitem{Litaize2018}
O.~Litaize~et al., ``{Influence of primary fragment excitation energy and spin
  distributions on fission observables},'' {\em EPJ Web of Conf.}, vol.~169,
  p.~00012, 2018.

\bibitem{Becvar1998}
F.~Be\v{c}v\'{a}\v{r}, ``Simulation of $\gamma$ cascades in complex nuclei with
  emphasis on assessment of uncertainties of cascade-related quantities,'' {\em
  Nucl. Instr. Meth. Phys. Res. A}, vol.~417, no.~2-3, pp.~434--449, 1998.

\bibitem{Fifrelin2016}
D.~Regnier~et al., ``An improved numerical method to compute neutron/gamma
  deexcitation cascades starting from a high spin state,'' {\em Comput. Phys.
  Commun.}, vol.~201, pp.~19--28, 2016.

\bibitem{Brown20181short}
D.~Brown, M.~Chadwick, R.~Capote, {\em et~al.}, ``{ENDF/B-VIII.0}: The
  {8$^{th}$} major release of the nuclear reaction data library with
  {CIELO}-project cross sections, new standards and thermal scattering data,''
  {\em Nuclear Data Sheets}, vol.~148, pp.~1 -- 142, 2018.
\newblock Special Issue on Nuclear Reaction Data.

\bibitem{jendl}
K.~K.~Shibata~et al., ``"jendl-4.0: A new library for nuclear science and
  engineering,'' {\em " J. Nucl. Sci. Technol.}, vol.~48, no.~1, pp.~1--30,
  2011.

\bibitem{Mughabghab}
S.~F. Mughabghab, {\em {Atlas of Neutron Resonances: Resonance Parameters and
  Thermal Cross Sections Z=1-100.}}
\newblock Elsevier Science, 2006.

\bibitem{RIPL3}
R.~Capote~et al., ``Reference input parameter library for calculation of
  nuclear reactions and nuclear data evaluations,'' {\em Nuclear Data Sheets},
  vol.~110, pp.~3107--3214, 2009.

\bibitem{Bethe1936}
H.~A.~Bethe, ``An attempt to calculate the number of energy levels of heavy
  nucleus,'' {\em Phys. Rev.}, vol.~50, no.~4, pp.~332--341, 1936.

\bibitem{GilbertCameron1965}
A.~Gilbert and A.~G.~W. Cameron, ``{A composite nuclear-level density formula
  with shell corrections},'' {\em Canadian Journal of Physics}, vol.~43,
  pp.~1446--1496, 1965.

\bibitem{Porter-Thomas1956}
E.~Porter and C.~R.~G.~Thomas, ``Fluctuations of nuclear reaction widths,''
  {\em Phys. Rev.}, vol.~104, p.~483, 1956.

\bibitem{Kibedi2008}
T.~{Kibédi et al.}, ``Evaluation of theoretical conversion coefficients using
  $\text{BrIcc}$,'' {\em Nucl. Instr. Meth. Phys. Res. A}, vol.~589,
  pp.~202--228, 2008.

\bibitem{KopeckyUhl}
J.~Kopecky and M.~Uhl, ``Test of gamma-ray strength functions in nuclear
  reaction model calculations,'' {\em Phys. Rev. C}, vol.~41, no.~5,
  pp.~1941--1955, 1990.

\bibitem{Brink1957}
D.~M.~Brink, ``Individual particle and collective aspects of the nuclear
  photoeffect,'' {\em Nucl. Phys.}, vol.~4, pp.~215--220, 1957.

\bibitem{GEANT4}
S.~Agostinelli~et al., ``{Geant4—a simulation toolkit},'' {\em Nucl. Instr.
  Meth. Phys. Res. A}, vol.~506, pp.~250--303, 2003.

\bibitem{GLG4sim_Gd}
{\url{http://neutrino.phys.ksu.edu/~GLG4sim/Gd.html}}.

\bibitem{SCHERZINGER201574}
J.~Scherzinger {\em et~al.}, ``Tagging fast neutrons from an
  $^{241}\text{Am}/^9\text{Be}$ source,'' {\em Applied Radiation and Isotopes},
  vol.~98, pp.~74 -- 79, 2015.

\bibitem{fifrelinDATA}
H.~Almazán {\em et~al.}, ``Data from: Improved \textsc{Stereo} simulation with
  a new gamma ray spectrum of excited gadolinium isotopes using
  \textsc{Fifrelin},'' May 2019, \\doi: 10.5281/zenodo.2653786.

\end{thebibliography}

\end{document}